\documentclass{aa}  

\usepackage{multirow}
\usepackage{graphicx}
\usepackage{txfonts}
\begin{document}

   \title{Detection of an ionized gas outflow in the extreme UV-luminous star-forming galaxy BOSS-EUVLG1 at z=2.47}

   \subtitle{}
  
  \titlerunning{Ionized gas outflow in BOSS-EUVLG1}
  
   \author{J.~\'Alvarez-M\'arquez\inst{1}
    \and R. Marques-Chaves\inst{1}
    \and L. Colina\inst{1,2}
    \and I. P\'{e}rez-Fournon\inst{3,4}
          }

   \institute{
    $^{1}$Centro de Astrobiolog\'ia (CSIC-INTA), Carretera de Ajalvir, 28850 Torrej\'on de Ardoz, Madrid, Spain; \email{javier.alvarez@cab.inta-csic.es} \\
    $^{2}$International Associate, Cosmic Dawn Center (DAWN)\\
    $^{3}$Instituto de Astrof\'\i sica de Canarias, C/V\'\i a L\'actea, s/n, E-38205 San Crist\'obal de La Laguna, Tenerife, Spain\\
    $^{4}$Universidad de La Laguna, Dpto. Astrof\'\i sica, E-38206 San Crist\'obal de La Laguna, Tenerife, Spain
             }

   \date{}

\abstract{BOSS-EUVLG1 is the most ultraviolet (UV) and Ly$\alpha$ luminous galaxy to be going through a very active
starburst phase detected thus far in the Universe. It is forming stars at a rate (SFR) of 955 $\pm$ 118 M$_{\odot}$ yr$^{-1}$. We report the detection of a broad H$\alpha$ component carrying 25\% of the total H$\alpha$ flux.
The broad H$\alpha$ line traces a fast and massive ionized gas outflow characterized by a total mass, $\log(M_{out}[M_{\odot}]),$ of 7.94 $\pm$ 0.15, along with an outflowing velocity (V$_{out}$) of 573 $\pm$ 151 km s$^{-1}$ and an outflowing mass rate ($\dot{M}_{out}$) of 44 $\pm$ 20 M$_{\odot}$ yr$^{-1}$. The presence of the outflow in BOSS-EUVLG1 is also supported by the identification of blueshifted UV absorption lines in low and high ionization states. The energy involved in the H$\alpha$ outflow can be explained by the ongoing star formation, without the need for an active galactic nucleus (AGN) to be included in the scenario. The derived low mass-loading factor ($\eta$= 0.05 $\pm$ 0.03) indicates that, although it is massive, this phase of the outflow cannot be relevant for the quenching of the star formation, namely, the negative feedback. In addition, only a small fraction ($\leq$ 15\%) of the ionized outflowing material with velocities above 372 km s$^{-1}$ has the capacity to escape the gravitational potential and to enrich the surrounding circumgalactic medium at distances above several tens of kpc. The ionized phase of the outflow does not carry sufficient mass or energy to play a relevant role  in the evolution of the host galaxy nor in the enrichment of the intergalactic medium. As predicted by some recent simulations, other phases of the outflow could be responsible for most of the outflow energy and mass in the form of hot X-ray emitting gas. The expected emission of the extended X-ray emitting halo associated with the outflow in BOSS-EUVLG1 and similar galaxies could be detected with the future ATHENA X-ray observatory, however, there are no methods at present that would assist in their spatial resolution. }


  \keywords{Galaxies: starburst -- galaxies: high-redshift -- galaxies: kinematics and dynamics -- galaxies: ISM -- intergalactic medium}

   \maketitle
%

\section{Introduction}

Cosmological simulations exploring the formation and evolution of galaxies in the early universe invoke outflows to explain the observed stellar mass function of galaxies. While outflows produced by luminous accreting black holes are favoured for the high-mass end, outflows associated with intense starbursts are claimed for the low-mass end  \citep[e.g.,][]{Somerville+15,Naab+17,Nelson+19,Mitchell+20a, Valentini+20}.  These outflows do not only control the growth of stellar mass in galaxies across redshifts, they are also responsible for the metal enrichment of the circum- and intergalactic medium  \citep{Kim+20}.

Observationally, outflows in galaxies are a universal phenomenon that have been detected at both low \citep[e.g.,][]{Arribas+14, Cicone+16, Concas+19} and high redshifts \citep[e.g.,][]{Freeman19, Forster-Schreiber+19, Swinbank19, Finnerty+20}, covering all types of galaxies, ranging from those featuring an active galactic nucleus (AGN) \citep[e.g.,][]{Harrison+14, King+15, Fiore+17, Harrison+18} to those featuring starbursts \citep[e.g.,][]{Heckman+1990, Veilleux+05, Veilleux+20}. The complex, intrinsic multi-phase nature (molecular, neutral, and ionized) of galactic outflows  has been traced at all wavelengths from the X-rays \citep{Strickland+04}, to ultraviolet \citep{Steidel+10}, optical \citep{Arribas+14, Cazzoli+16, Cicone+16, Concas+19}, near-infrared \citep{Hill+14, Emonts+17}, far-infrared \citep{Veilleux+13}, and millimeter wavelengths \citep{Cicone+14, Pereira-Santaella+18, Spilker+20a, Spilker+20b, Veilleux+20}.

Most of the studies of ionized outflows at high-$z$ have focused on galaxies located on the main sequence (MS) of star-forming galaxies (SFG), namely, galaxies with average star formation rates (SFRs) of several to a few tens of M$_{\odot}$ yr$^{-1}$ \citep{Forster-Schreiber+19, Freeman19, Swinbank19}. Ionized gas outflows appear to be frequent at the peak of the cosmic star formation history (i.e. z$\sim$1-3). As concluded in the recent KMOS$^{3D}$ H$\alpha$ survey \citep{Forster-Schreiber+19}, about a third of the non-AGN dominated SFGs show evidence of ionized gas outflows with masses (M$_{out}$) of 10$^7$ to 10$^8$ M$_{\odot}$, outflowing mass rates ($\dot{M}_{out}$) of few tens M$_{\odot}$ yr$^{-1}$, and low mass-loading factors ($\eta = \dot{M}_{out}$/SFR) of $\sim$0.1-0.2. Similar mass outflowing rates and mass-loading factors have been measured in MS star-forming galaxies at z$\sim$1 \citep{Swinbank19}. These studies concluded that the warm ionized phase of the outflow is not sufficient to explain low-mass galaxy formation efficiency, additionally asserting that cold molecular and hot X-ray emitting phases are expected to dominate the mass, energy, and momentum of the outflows as observed in low-z galaxies \citep[e.g.,][]{Veilleux+20, Strickland+04}. 

Studies of outflows in high-z galaxies with extreme star formation, that is, with SFRs of several hundreds to thousand M$_{\odot}$ yr$^{-1}$, are far more limited. A recent study of massive, dusty star-forming galaxies (DSFG) at z$>$4  reported the detection of massive molecular outflows in a large fraction (73$\%$) of the galaxies \citep{Spilker+20a}. These outflows are characterized by high velocities (V$_{max}$ $\sim 430 - 1200$ km s$^{–1}$), mass outflowing rates of several M$_{\odot}$ yr$^{-1}$, and mass-loading factors just below one \citep{Spilker+20b}. However, the presence of the (partially) ionized outflowing component was not detected in those sources with a high quality [CII]158$\mu$m line and with clear molecular outflow, as traced by the OH119$\mu$m absorption line. High-velocity outflows have also been detected in z > 2 Lyman break galaxies (LBGs, \citealt{Steidel+10, Sugahara+19}). These LBGs show blueshifted low- and high-ionization ultraviolet (UV) interstellar medium (ISM) absorption lines tracing outflows with maximum velocities of about 800 km s$^{-1}$. Therefore, high-velocity outflows appear to be common in both infrared (i.e., dusty) and ultraviolet (i.e., dust-poor) luminous star-forming galaxies at high redshifts. However, the multi-phase structure and their impact on the formation and evolution of galaxies is far from clear. Detailed multi-wavelength studies of individual prototypes, or small samples of galaxies, are required to quantify and establish the relevance of each of the main outflow components (cold dense molecular, warm (partially-) ionized, or hot diffuse X-rays) in the stellar buildup of galaxies and in the metal enrichment of the intergalactic medium (IGM).

Very recently, an extremely UV luminous galaxy, BOSS-EUVLG1, at a redshift of 2.47, was discovered in the study of \citeauthor{Marques-Chaves20} (2020; hereafter, MC20 ) within the extended Baryon Oscillation Spectroscopic Survey \citep[][]{abolfathi2018}. BOSS-EUVLG1 is a compact (size $\leq$ 1.2 kpc) and young ($\sim$4-5 Myr) starburst forming stars at a rate of  $\sim$ 1000 M$_{\odot}$ yr$^{-1}$, with low metallicity and dust content (12+log(O/H)=8.13$\pm$0.19, E(B-V)$\simeq$0.07). BOSS-EUVLG1 is the most UV-luminous galaxy identified so far in the universe and can be considered as the prototype of a new and unique type of extremely UV luminous galaxies (EUVLGs) characterized as having extremely high UV continuum and Ly$\alpha$ luminosities ($M_{\rm UV} < -23.5$; log($L_{\rm Ly\alpha} [\rm erg~s^{-1}$]) > 43.5), and unobscured SFR above 450 M$_{\odot}$ yr$^{-1}$. In addition, BOSS-EUVLG1 has a specific star formation (sSFR) $\sim$ 106 $\pm$ 15 Gyr$^{-1}$, a factor 30 times higher than 10$^{10}$ M$_{\odot}$ main-sequence galaxies at redshifts 2-2.5 \citep{whitaker+15} but in the range of the H$\alpha$-excess galaxies at redshifts $\sim$ 4 to 5 that were recently identified in the {\it Spitzer} UltraVISTA ultra-deep stripes survey (SMUVS survey; \citealt{Caputi+17}).  An important fraction (15\%) of the SMUVS galaxies in the 9.4 $\leq$ log(M$_{stellar}[M_{\odot}]) \leq$ 11.0 mass range appear to be in a starburst phase with SFR of hundreds to thousands M$_{\odot}$ yr$^{-1}$ and sSFR well above 24 Gyr$^{-1}$, and up to 120 Gyr$^{-1}$. The SMUVS galaxies in the starburst phase account for more than 50\% of the cosmic SFR density at redshifts $\sim$ 3.9–4.9. Therefore, extreme UV luminous galaxies such as BOSS-EUVLG1 could represent the short lived phase of this class of H$\alpha$-excess galaxies, and could be important contributors to the SFR density at intermediate and high redshifts. BOSS-EUVLG1, therefore, brings on the opportunity to investigate the presence and properties of ionized gas outflows for the first time in this class of extreme star-forming galaxies.

In this article, we report the detection of the ionized phase of the outflow in BOSS-EUVLG1 using the H$\alpha$ emission line. Section \ref{Sect:observations} describes the observations and calibrations of the H$\alpha$ spectrum. Section \ref{Sect:data_analysis} presents the analysis of H$\alpha$ emission and rest-frame UV absorption lines. The results of the analysis and interpretation as outflowing material are presented in Section \ref{Sect:results}, including the overall kinematic and mass ($\S$\ref{Sect:kinematics}), the outflowing mass rate and mass-loading factor ($\S$\ref{Sect:massrate}), the nature of the energy source ($\S$\ref{Sect:energy}), and the mass escape fraction into the intergalactic medium ($\S$\ref{Sect:escape}). The results are discussed in Section \ref{Sect:Discussion}, focusing in the comparison with low- and high-z SFGs ($\S$\ref{Section:Disc1}), the multi-phase nature of the outflows ($\S$\ref{multiphase}), and the possibility of detecting these outflows with the future X-ray observatory, {\it ATHENA} ($\S$\ref{ATHENA}). Finally, Section \ref{Sect:summary} present the summary and outlook. We adopted $H_{0} = 67.7$ km s$^{-1}$ Mpc$^{-1}$ and $\Omega_{\rm m} = 0.307$ \citep{Cosmology2016} as cosmological parameters and the Salpeter initial mass function (IMF; \citealt{Salpeter+55}) throughout this paper. A factor of 1.66 \citep{Newman+12} is used to convert the SFRs and stellar masses (M$_{stellar}$) from a \cite{Chabrier2003} to a \cite{Salpeter+55} IMF when using some published data.

\section{Observations}\label{Sect:observations}

Near-infrared (Near-IR) long-slit spectroscopy of BOSS-EUVLG1 was obtained with the {\it Espectr\'ografo Multiobjeto Infra-Rojo} (EMIR) instrument on the 10.4m Gran Telescopio Canarias (GTC). Observations were taken in two service mode nights -- 2020 February 10 and May 8 -- in bright Moon and subarcsec seeing ($\simeq$0.8$^{\prime \prime}$ full width half maximum) conditions, as part of the GTC program GTCMULTIPLE2E-19A (PI: R.~Marques-Chaves). The observations used the $K$ grism with a dispersion of 1.71~\AA~pix$^{-1}$ and a 0.8$^{\prime \prime}$-wide long-slit, providing a spectral resolution of 82~km~s$^{-1}$ at $\sim$2.25$\mu$m. The slit was centered on a bright ($K=17.43\pm0.02$, AB) reference star (SDSS J122039.27+084225.0, $\simeq25^{\prime \prime}$ SW of BOSS-EUVLG1) and oriented so as to encompass BOSS-EUVLG1, with a sky position angle of PA=58.8$^{\circ}$ (north toward east).
The total exposure time on-source was 6400~s, split into 40 individual exposures of 160~s each, using an ABBA nodding pattern with 6$^{\prime \prime}$ offset along the slit. 

Data reduction was performed using the EMIR pipeline  ({\sc PyEmir})\footnote{\url{https://pyemir.readthedocs.io/en/latest/index.html}}.  Near-IR 2D spectra were flux calibrated using the HIP~59174 telluric standard star observed in the second night. To account for slit-losses, the fluxes of BOSS-EUVLG1 and the reference star were matched to those obtained from photometry. The 1D spectrum of H$\alpha$ was extracted applying an optimal extraction algorithm \citep{Horne86} in an aperture of 3$^{\prime \prime}$. Figure \ref{fig:Halpha_fit} shows the final 2D calibrated and 1D extracted spectrum of BOSS-EUVLG1.

\section{Data analysis}\label{Sect:data_analysis}

\begin{figure}[htb!]
  \centering
    \includegraphics[width=\hsize]{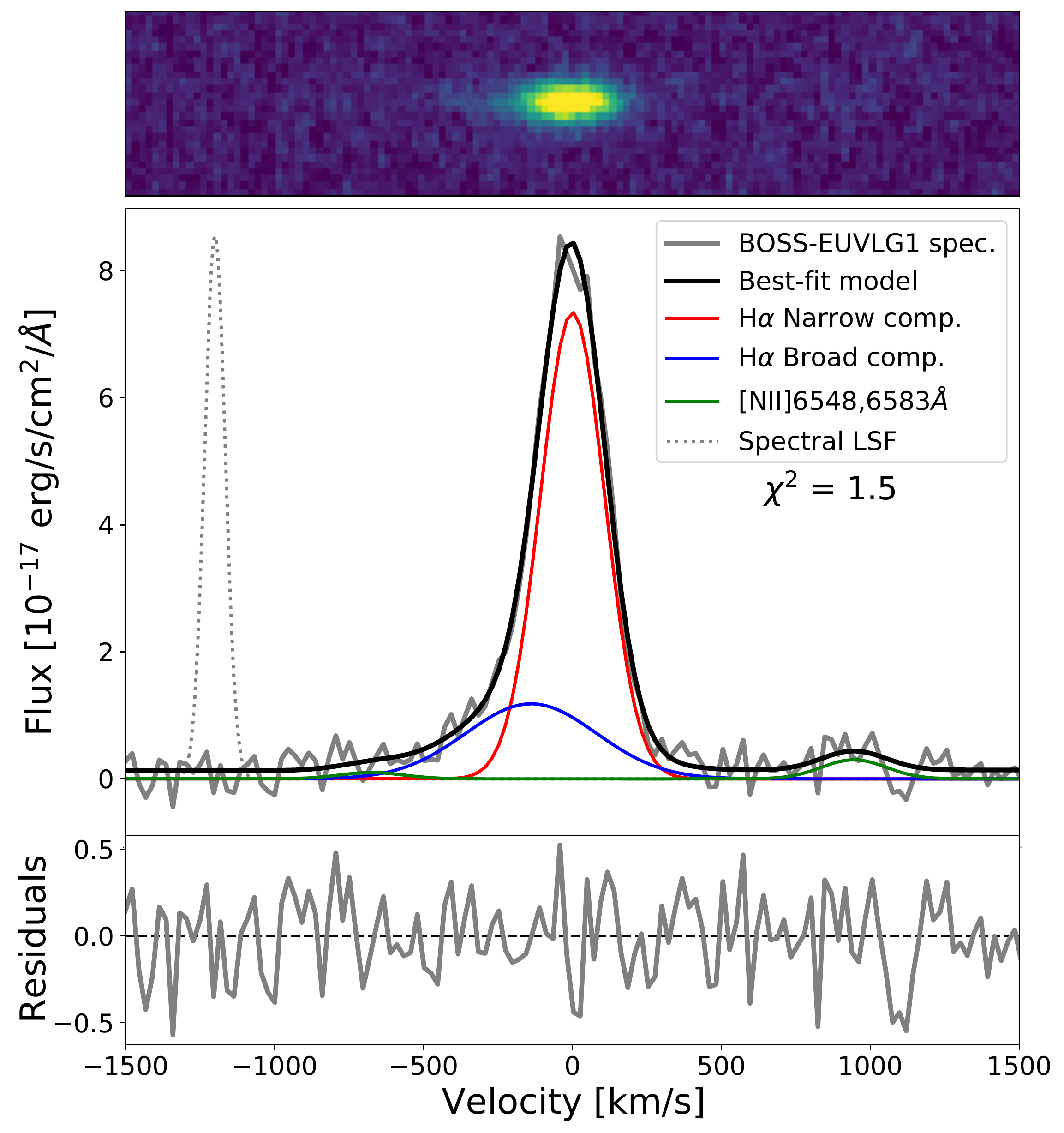}
\caption{Observed H$\alpha$ emission of BOSS-EUVLG1. Upper panel shows the near-IR 2D calibrated spectrum centered on the H$\alpha$ emission. Bottom panel illustrates the observed H$\alpha$ emission line profile of BOSS-EUVLG1, together with the two-components Gaussian fit and its residuals. Gray line: 1D observed  H$\alpha$ spectrum. Black line: Best two-component Gaussian H$\alpha$ fitted model. Red and blue lines: Narrow and broad  H$\alpha$ components. Green line: [N~{II}]6548,6583$\AA$ lines. Gray dotted line: Illustration of observed spectral line-spread function, derived from a sky line at 2.25$\mu$m, normalized to the peak of the H$\alpha$ emission, and centered within a velocity of -1200km/s. The $\chi^{2}$, calculated from -1000 to 750 km s$^{-1}$, is also included.}\label{fig:Halpha_fit}
\end{figure}

The H$\alpha$ emission of BOSS-EUVLG1 exhibits an asymmetric line profile characterized by a main component with a blue-shifted wing (Figure \ref{fig:Halpha_fit}). We implement a two-component Gaussian fit, using a Python version of the MPFIT routine \citep{Markwardt2009}\footnote{MPFIT Python version used: \url{https://github.com/segasai/astrolibpy/tree/master/mpfit}}, to characterize the H$\alpha$ emission.  The [N~{II}]6548,6583$\AA$ doublet and a continuum emission, together with the instrumental broadening, are also included in the fit. Figure \ref{fig:Halpha_fit} shows the results of the two-components fit with the narrow and a blue-shifted broad emission of H$\alpha$. Additionally, we test the possibility to fit the profile with just one Gaussian component but it fails to reproduce its asymmetry, yielding to a $\chi^{2}$ 50\% larger than using the two-component fit. The errors on the parameters are estimated using Monte Carlo. The noise of the spectrum is measured as the root mean square (rms) of the continuum surrounding the H$\alpha$+[NII] lines. This noise is used to generate new spectra (3000), where a random Gaussian noise with a sigma equal to the rms is added to the original spectrum before the lines are fitted again.

The H$\alpha$ narrow component has a flux (F$_{N}$) of $1.43 \pm 0.16\times10^{-15}$ ~erg~s$^{-1}$~cm$^{-2}$ and a full width at half maximum (FWHM$_{N}$) of $241 \pm 11$~kms$^{-1}$. The broad component is characterized by an H$\alpha$ flux (F$_{B}$) of $4.9 \pm 1.7\times10^{-16}$ ~erg~s$^{-1}$~cm$^{-2}$, a width  (FWHM$_{B}$) of $511 \pm 145$~kms$^{-1}$, and a velocity offset (V$_{peak}$) of $-139 \pm 87$~kms$^{-1}$ (i.e., blueshift) with respect to the narrow component. In addition, the [N~{II}]6583$\AA$ emission appears as a weak line with a measured flux of $(5.9 \pm 1.5)\times10^{-17}$~erg~s$^{-1}$~cm$^{-2}$ for the best fit. These results are consistent, within the uncertainties, with the slight asymmetries in the low-resolution spectrum of the [O~{\sc iii}]~$4959,5007\AA$ lines (R$\sim$735, MC20). Higher spectral resolution is required to constrain the properties of the outflow traced by the high ionization [O~{\sc iii}]$~4959,5007\AA$ lines.

Evidence for outflowing material in BOSS-EUVLG1 comes from the analysis of the UV rest-frame absorption line spectrum taken with the GTC (see MC20 for a complete description of the observed spectrum). We use several UV ISM absorption lines in a variety of ionization states, from partial neutral gas (Si~{\sc ii}$~1260\AA$, C~{\sc ii}$~1334\AA$, and Si~{\sc ii}$~1526\AA$) to highly ionized species (Si~{\sc iv}$~1393,1402\AA$). The rest-frame UV spectrum of BOSS-EUVLG1 is first normalized using spectral windows that are free of absorption and emission features identified by \cite{Rix2004}. Next, we derive the average ISM absorption line profiles by performing a mean of the individual absorption lines in the velocity plane. Left panel of Figure \ref{fig:HalphaUV} shows the resulting average profiles using low- and high-ionization UV ISM absorption lines and the comparison with the H$\alpha$ line. As seen in this figure, the centroids of low- and high-ionization ISM absorption lines are clearly blueshifted with respect to the systemic redshift by $\Delta v_{\rm low-ion} = -305 \pm 75$~km~s$^{-1}$ and $\Delta v_{\rm high-ion} = -415 \pm 75$~km~s$^{-1}$, respectively. The errors account for the uncertainties of the spectral bin size and the standard deviation of the centroid of each individual ISM line. Due to the low spectral resolution of the rest-frame UV spectrum ($R\sim700$), these lines appear almost unresolved so that the detailed characterization of the outflowing gas from ISM absorption lines (i.e., $\dot{M}_{out}$, $V_{\rm max}$, column density, covering fraction, etc.), is not possible as it requires a better spectral resolution. However, they do present further evidence for the existence of outflows in this galaxy and, therefore, lend their support to the interpretation of the broad H$\alpha$ component as outflowing material.

\section{Results}\label{Sect:results}
\subsection{Kinematics and mass of the ionized outflow}\label{Sect:kinematics}
The narrow component of H$\alpha$ is assigned to the emission in the host galaxy while the broad component is identified with outflowing ionized gas. The main properties of this outflowing material (i.e., outflowing mass, velocity of the bulk of the outflow and its maximal velocity)  are derived from the flux, offset velocity, and line profile of the broad H$\alpha$ line component, assuming the electron density is known. For BOSS-EUVLG1, the electron density is derived from the  low-ionization [O~{\sc ii}]$~3729,3727\AA$ ratio (0.68 $\pm$0.21, MC20) giving a density of 1500 cm$^{-3}$ \citep{Kewley19}. This value provides the mean electron density of the ionized gas in the galaxy. Since no direct measurement of the electron density in the outflow is available within our data, this value of the density will also be assumed for outflowing material in all the subsequent analysis and derivation of the physical quantities. This is a valid assumption. The median electron density of the outflows traced by the low ionization [SII]6717,6731$\AA$ lines in a large sample of low-z U/LIRGs is a factor of 1.23 larger than the density of the ionized gas in these galaxies \citep{Arribas+14}. 

The peak velocity (V$_{peak}$) of the broad H$\alpha$ component is blueshifted by $-$139 $\pm$ 87 km s$^{-1}$. High-spatial resolution zoom-in simulations of the ionized gas velocity field traced by the H$\alpha$ emission line in $z \sim$ 2 non-interacting galaxies predict  the presence of a broad H$\alpha$ line component in two-third of the simulated galaxies and with blueshifted velocities of 110 $\pm$ 20 km s$^{-1}$ \citep{Ceverino+16}. This result is explained as a direct consequence of the extended structure of the outflowing material, coming from the near-side of the galaxy, above the plane of the galaxy \citep{Ceverino+16}. These simulations cover galaxies with stellar masses 1-4 $\times$ 10$^{10}$ M$_{\odot}$, similar to that of BOSS-EUVLG1, but with SFR $\sim$ 20-60 M$_{\odot}$ yr$^{-1}$, namely, factors of 16-50 lower than the SFR derived for BOSS-EUVLG1. A similar spatial and velocity structure can be expected in BOSS-EUVLG1 where, due to its higher SFR, the outflowing material could be launched to distances further out, and up to several kpc above the plane of the galaxy.

The observed H$\alpha$ outflowing velocity is substantially lower than that measured using the UV absorption lines ($\Delta v_{\rm low-ion} = -305 \pm 75$~km~s$^{-1}$ and $\Delta v_{\rm high-ion} = -415 \pm 75$~km~s$^{-1}$), but it is in agreement with the typical blueshifted velocity observed in Lyman Break Galaxies at intermediate redshifts ($-$164 $\pm$ 16 km s$^{-1}$,  \citealt{Steidel+10}). Such differences are expected because the outflowing gas is stratified, that is, regions of different ionization state (neutral versus ionized gas, low- versus high-ionization ISM) show different outflowing velocities \citep[e.g.,][]{Erb15, Rupke+18, delacruz2020}. 

 \begin{figure*}[htb!]
  \centering
    \includegraphics[width=\hsize]{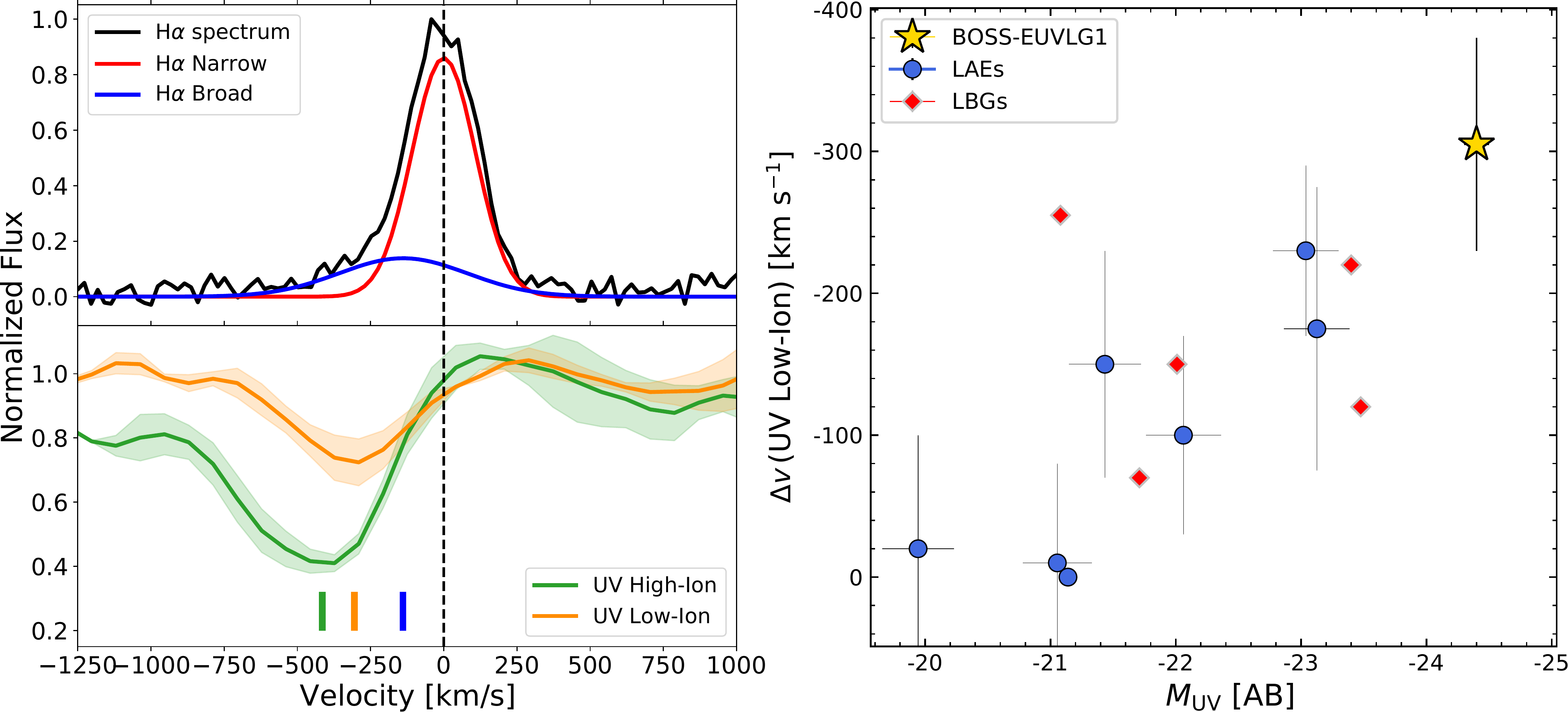}
\caption{View of the UV absorption lines and the UV peak velocity of BOSS-EUVLG1. Left panel: Comparison between the H$\alpha$ emission and the UV ISM absorption lines. Upper left panel: Normalized H$\alpha$ spectrum (black line), together with the narrow (red line) and broad (blue line) H$\alpha$ components. Bottom left panel: UV ISM absorption lines for two different ionization states. Orange line: Combination of UV ISM absorption lines of partially neutral elements (Si~{\sc ii}$~1260\AA$, C~{\sc ii}$~1334\AA$, and Si~{\sc ii}$~1526\AA$). Green line: Combination of UV ISM absorption lines of high ionization elements (Si~{\sc iv}$~1393,1402\AA$). Both lines are normalized in a spectral windows clean of absorption and emission features. The peak velocities of both UV absorption ionization states and the broad H$\alpha$ component are shown as color-coded vertical lines. Right panel: Peak velocity of the UV low-ionization ISM absorption lines as a function of the UV absolute magnitude. Yellow star: BOSS-EUVLG1. Blue dots: Lyman alpha emitters \citep{patricio2016, marques2020}. Red squares: Lyman break galaxies \citep{pettini2000, quider2009, quider2010, des2010, marques2018}.}\label{fig:HalphaUV}
\end{figure*}

The kinematics of the H$\alpha$ outflow is characterized by the peak velocity offset and the outflow velocity ($V_{out}= V_{peak} + 2 \times \sigma_{B}$). For a FWHM$_B$ of 511 $\pm$ 145 km s$^{-1}$ and a V$_{peak}$ of -139 $\pm$ 87 km s$^{-1}$, the V$_{out}$ correspond to 573 $\pm$ 151 km s$^{-1}$. Following \cite{Colina91}, the total outflowing ionized mass (M$_{out} \propto$ L(H$\alpha$)/n$_{e}$) corresponds to $\log(M_{out}[M_{\odot}]) = 7.94 \pm 0.15$, derived from the extinction corrected (E(B-V)=0.07, MC20) broad H$\alpha$ luminosity. A more conservative estimate of the total outflowing ionized mass can be derived by assuming only the ionized gas with velocities that are well above the expected dispersion or rotational velocities derived from the narrow H$\alpha$ component. Considering velocities only above 241 km s$^{-1}$ (FWHM$_{N}$) and the offset velocity of the outflowing component, the lower limit to the outflowing mass would correspond to about a third of the total mass. Therefore, the outflowing ionized mass would be in the mass range of 7.42 < $\log(M_{out}[M_{\odot}]) < 7.94$.

\subsection{Outflowing mass rate and mass-loading factor}\label{Sect:massrate}
The outflowing ionized mass rate is given as $K \times M_{out} \times V_{out} \times size^{-1}$, where $K$ is a factor that depends on the geometry, with values between 1 and 3 for an elongated and spherical filled outflow, respectively  (\citealt{Lutz+19}). Here, we adopt a value of one as the spatially resolved outflows in nearby galaxies support in general elongated geometries \citep{Pereira-Santaella+18} and the limited spatial resolution here does not provide information on the geometry of the outflow. For a V$_{out}$ of 573$\pm$ 151 km s$^{-1}$ and an outflow size comparable to the UV-continuum source  (i.e., 1.2 kpc, MC20), the outflowing mass rate corresponds to 44 $\pm$ 20 M$_{\odot}$ yr$^{-1}$.

The SFR derived from the observed H$\alpha$ flux corrected by the internal extinction (E(B-V)=0.07, MC20) corresponds to $955 \pm 118$ M$_{\odot}$  yr$^{-1}$ for a Salpeter IMF \citep{Salpeter+55} and a SFR = 7.9$\times 10^{-42} \times L(H\alpha)$ \citep{Kennicutt1998}. The corresponding mass-loading factor ($\eta$= $\dot{M}_{out}$/SFR) is 0.05 $\pm$ 0.03.  This low value implies that while the amount of ionized gas in the outflow is significant,  it does not represent a substantial amount relative to the ongoing star formation process that is taking place. Therefore, the ionized warm phase of the outflow, traced by the H$\alpha$ emission line,  appears irrelevant in the quenching of the extreme on-going star formation in BOSS-EUVLG1. However, other phases could serve as the dominant phases of the outflow, carrying most of its mass and energy,  therefore acting as the major contributors to the removal of gas and quenching of star formation (see discussion in $\S$ \ref{multiphase}).

\subsection{Energy source of the outflow}\label{Sect:energy}
 
For young starbursts, the rates of outflowing mass, kinetic energy, and momentum liberated by supernovae explosions and stellar winds are a function of the SFR. Following \cite{Veilleux+05}, these  quantities correspond to 248 M$_{\odot}$ yr$^{-1}$, $6.7 \pm 0.8 \times 10^{44}$ erg s$^{-1}$, and $4.8 \pm 0.6 \times 10^{36}$ dynes  for a SFR of 955 M$_{\odot}$ yr$^{-1}$, respectively. The measured rate of kinetic energy ($0.5 \times \dot{M}_{out} \times V_{out}^2$) and momentum ($\dot{M}_{out} \times V_{out}$) of the H$\alpha$ outflow corresponds to 4.6 $\pm$ 3.2 $\times 10^{42}$ erg s$^{-1}$ and 1.6 $\pm 0.8  \times 10^{35}$ dynes, respectively. The ionized outflow represents in BOSS-EUVLG1 a minor fraction of the  outflowing mass (18\%), kinetic energy (0.7\%), and momentum (3.3\%) rates liberated by the starburst. There is therefore no need to invoke the presence of an AGN in the nucleus to explain the energetics of the warm ionized outflow. However, unless the starburst in BOSS-EUVLG1 is extremely inefficient in converting the available energy into outflows in the surrounding interstellar medium, contributors to the outflow could still be present and locked in other phases such as in the hot, X-ray emitting gas (see $\S$\ref{multiphase} for further discussion).

 \begin{figure*}[htb!]
  \centering
    \includegraphics[width=\hsize]{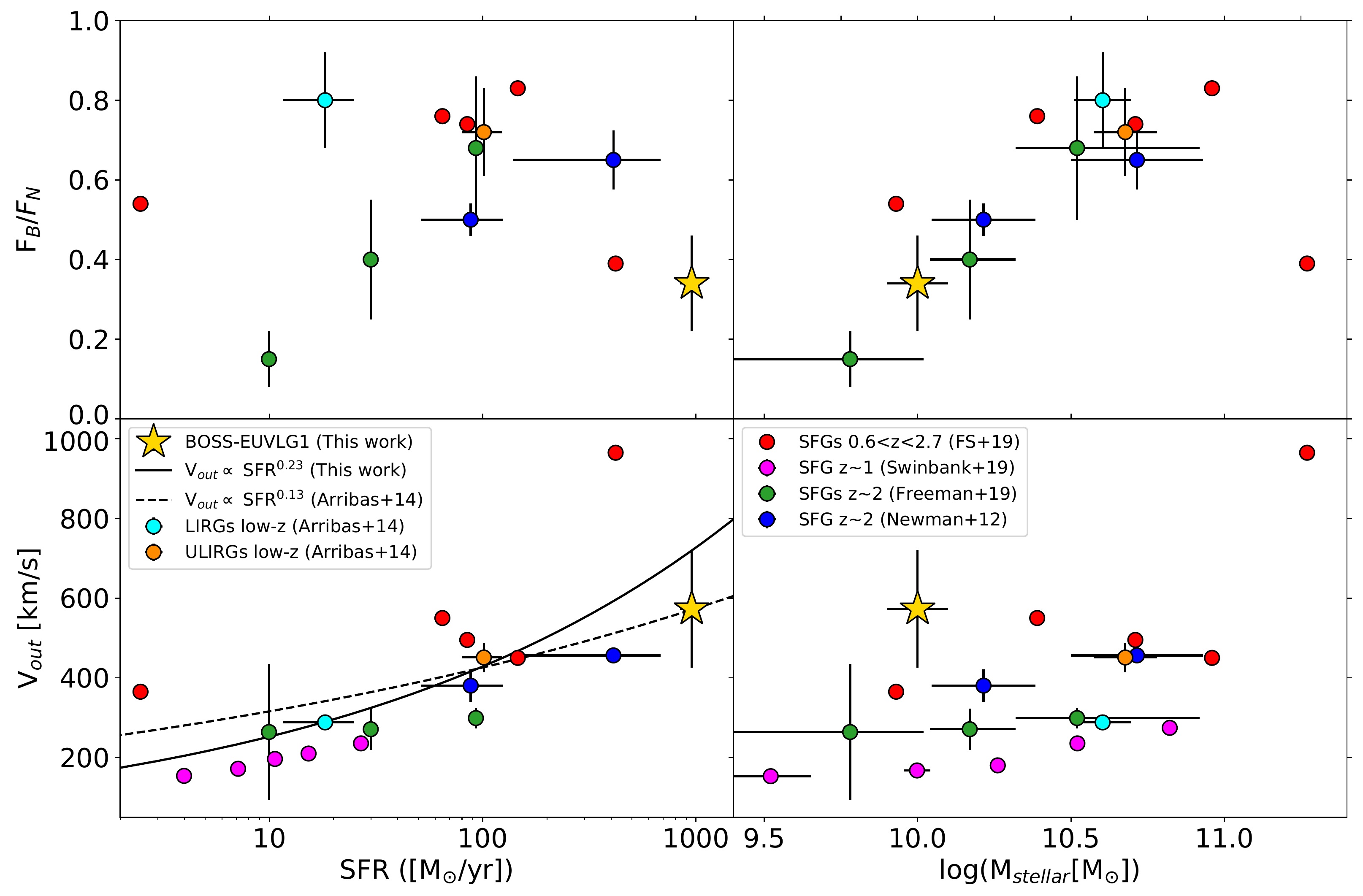}
\caption{Observational properties of the ionized outflow in BOSS-EUVLG1 (yellow star) together with samples of star-forming galaxies (SFGs) at intermediate redshifts including 0.6<z<2.7 SFGs (red dots) from the KMOS$^{3D}$ survey  (\citealt{Forster-Schreiber+19}, values of V$_{out}$ are a private communication), SFGs at z$\sim$1 (magenta dots) from KMOS observations \citep{Swinbank19}, SFGs at z$\sim$2 (green dots) from the MOSDEF survey \cite{Freeman19}, and SFGs at $z\sim$2 (blue dots) from the SINS and zC-SINF surveys \citep{Newman+12}. Low-z (ultra-)luminous galaxies (cyan and orange dots) are also represented \citep{Arribas+14}. Upper panels present the flux ratio of the broad to narrow H$\alpha$ line components (F$_{B}$/F$_{N}$) as a function of the SFR (left) and M$_{stellar}$ (right). Bottom panels show the outflow velocity as a function of the SFR (left) and M$_{stellar}$ (right). The continuum line in the V$_{out}$ to SFR relation defines the best fit ($V_{out} = 148 \times SFR^{0.23}$) to the BOSS-EUVLG1, low- and intermediate-z data points. The dashed line represents the V$_{out}$ to H$\alpha$-derived SFR relation derived for the low-z sample of U/LIRGs without AGNs \citep{Arribas+14}. }\label{fig:results}
\end{figure*}

\subsection{Ionized mass escape fraction and IGM metal enrichment}\label{Sect:escape}

 Independent of the energy source, gas outflows in galaxies are relevant, not only as potential mechanisms of star formation quenching, but also as one of the main mechanisms for expelling gas from galaxies into the intergalactic medium, with subsequent metal enrichment (\citealt{Somerville+15}). An estimate of the fraction of the outflowing material in BOSS-EUVLG1 that is capable of escaping the gravitational potential of the galaxy and enrich the IGM can be derived assuming only the outflowing material with velocities above the escape velocity. For a galaxy represented by an isothermal sphere truncated to a maximum radius (R$_{max}$), the escape velocity for a dynamical mass (M$_{dyn}$) within a radius (R) is given as $V_{esc} = \sqrt{2 \times M_{dyn} \times G \times (1+ln[R_{max}/R]) / (3 \times R)}$ (e.g., \citealt{Arribas+14} and references therein). BOSS-EUVLG1 has a dynamical mass of 1.8 $\times 10^{10}$ M$_{\odot}$, derived for a R of 1.2 kpc and velocity dispersion from the narrow H$\alpha$ component of 102 km s$^{-1}$ \citep{Bellocchi+13}. For a R$_{max}$ of 10~kpc, the escape velocity corresponds to 372 km s$^{-1}$. 
 
 The velocity of the outflow (V$_{out}$= 573 km s$^{-1}$) is 1.5 times larger than the escape velocity. Thus, only a small fraction  of the total ionized outflowing material (15\%, i.e., $\sim$ 1.3 $\times 10^{7}$ M$_{\odot}$), with velocities well above V$_{esc}$, would potentially be able to escape and enrich the surrounding IGM at distances larger than 10~kpc. Most of the outflowing material would not escape the gravity of the host galaxy and would be likely to fall back into the system, providing the raw material for subsequent delayed star formation episodes, as in low-z infrared luminous star-forming galaxies \citep{Arribas+14,Ceverino+16}. We note, however, that H$\alpha$ only traces the warm (10$^4$ K) ionized gas component of the outflow. In fact, the peak velocities of the blueshifted low- and high-ionization UV ISM absorption lines ($\Delta v_{\rm low-ion} \sim -305$~km~s$^{-1}$ and $\Delta v_{\rm high-ion} \sim -415$~km~s$^{-1}$) are factors of 2 and 3 larger than the corresponding H$\alpha$ (V$_{peak}=-139 \pm 85$ km s$^{-1}$). This already demonstrates  the existence of outflowing material that is rich in metals, with velocities higher than those of the H$\alpha$ emitting gas -- velocities that are high enough to escape into the IGM and contribute to its metal enrichment.  
 
\section{Discussion}\label{Sect:Discussion}

\subsection{Ionized Outflows in BOSS-EUVLG1, low- and intermediate-z star-forming galaxies}\label{Section:Disc1}

BOSS-EUVLG1 is an extreme starburst galaxy with a star-forming rate of $955 \pm 118$ M$_{\odot}$ yr$^{-1}$, which can be considered an ultraluminous galaxy based on its UV luminosity (i.e., $L_{1500\AA} \sim 10^{12} L_{\odot}$). Its SFR is similar to those measured in high$-z$ dust-enshrouded extreme starbursts \citep{Reuter+20}, and about 10 to 100 times higher than that of massive star-forming galaxies at intermediate redshifts \citep{Forster-Schreiber+19, Swinbank19, Freeman19, Newman+12}, and low$-z$ luminous and ultraluminous infrared galaxies (U/LIRGs, \citealt{Arribas+14}). Moreover, BOSS-EUVLG1 is deficient in terms of dust and is underluminous in the infrared (log($L_{IR}$[L$_{\odot}])<$ 10.91), yielding a IR/UV luminosity ratio of log($L_{\rm IR}/L_{\rm UV}$)<$-$1.2. The low extinction and metallicity values measured in BOSS-EUVLG1 suggest that this galaxy is likely in an early and short-lived evolutionary stage as it continues to build its stellar mass, providing metal and dust enrichment, before evolving into the infrared luminous dust-enshrouded starburst phase (MC20).

 The observed properties of the outflow in BOSS-EUVLG1 are presented in Figure \ref{fig:results} together with those of low- and intermediate-z star-forming galaxies using the same tracer (H$\alpha$). The ratio of the H$\alpha$ broad- to narrow-component flux (F$_B$/F$_N$) in BOSS-EUVLG1 is $0.34 \pm 0.12$. This value is lower than the ratios measured in low- and intermediate-z star-forming galaxies where the broad line component carries a larger fraction of the flux with typical F$_B$/F$_N$ values in the 0.4 to 0.8 range, independent of the SFR albeit with a large scatter (upper left panel of Figure \ref{fig:results}). However, the F$_B$/F$_N$ ratio as a function of the stellar mass follows the expected trend for star-forming galaxies (upper right panel of Figure  \ref{fig:results}). These results clearly indicate that the ionized outflow component in BOSS-EUVLG1 is low for its specific star formation rate (sSFR $= 106 \pm 15$ Gyr$^{-1}$, MC20). BOSS-EUVLG1 is an extreme starburst galaxy with sSFR about 30-50 times higher than that of Main Sequence star-forming galaxies at z$\sim$2 \citep{Rodighiero2014}. There is no clear explanation for the low contribution of the outflowing component to the total H$\alpha$ emission. It appears as if the strength of the outflow is dominated by the stellar mass of the system -- and not by the strength of the star formation as given by the SFR. This is somehow unexpected given that in a starburst-dominated outflow, the mass, momentum, and energy rates of the outflow are directly related to the total star formation rate \citep{Heckman+1990,Veilleux+05}. In addition, outflows in z$\sim$2 star-forming galaxies are characterized by high F$_B$/F$_N$ values close to 1 for all galaxies with SFR surface brightness above 1 M$_{\odot}$ yr$^{-1}$ kpc$^{-2}$ and low values of about 0.2 for the galaxies below the SFR brightness threshold, independent of the stellar mass \citep{Newman+12}. According to its sSFR, SFR, and upper limit on the size of the H$\alpha$ emission, BOSS-EUVLG1 would be classified in the category of galaxies with strong outflows characterized by a F$_B$/F$_N$ ratio of about 1. However, the measured value of 0.34 is closer to that of weak outflows in starbursts with a low-SFR surface brightness.  Explanations of the low F$_B$/F$_N$ value in BOSS-EUVLG1 could invoke a qualitatively different multi-phase structure and composition of the interstellar medium and outflows in IR-luminous, namely, dustier, more molecular, and metal-rich;  and in UV-luminous starbursts such as BOSS-EUVLG1, namely, dust-, molecular-, and metal-poor (see discussion in $\S$\ref{multiphase} and $\S$\ref{ATHENA}). The {\it Spitzer} UltraVISTA ultra-deep stripes survey (SMUVS survey; \citealt{Caputi+17}) has identified a number IR-luminous, H$\alpha$-excess galaxies at redshifts of $\sim$ 4 to 5 with extreme sSFR. An important fraction (15\%) of the SMUVS galaxies in the 9.4 $\leq$ log(M$_{stellar}[M_{\odot}]) \leq$ 11.0 mass range appear to be in a starburst phase with SFR of hundreds to thousands M$_{\odot}$ yr$^{-1}$ and sSFR well above 24 Gyr$^{-1}$, and up to 120 Gyr$^{-1}$. Future H$\alpha$ spectroscopy of EUVLGs and SMUVS galaxies in the starburst phase from the ground and with JWST (for $z>3$) are required to confirm the strength of outflows in extreme starbursts (i.e., very large sSFR)  and investigate it in depth to identify the potential differences between the classes of IR- and UV-luminous. 
 
 The outflow velocity in BOSS-EUVlG1 is in agreement with the expected value as extrapolated from previous V$_{out}$ $\propto$ SFR$^{\alpha}$ relations derived in low-z U/LIRGs ($\alpha$= 0.21-0.24,  \citealt{Arribas+14} and references therein). This ratio is also consistent with the dependence observed in the ionized outflows of intermediate-z star-forming galaxies (lower-left panel of Figure \ref{fig:results}). The best fit, which includes BOSS-EUVLG1 as well as low- and intermediate-z galaxies (i.e., covering a factor of 400 in SFR) gives a slope ($\alpha$) of 0.23, which is consistent with the predicted dependence with the bolometric luminosity of the starburst (V$_{out}$ $\propto$ L$_{bol}^{0.25}$, \citealt{Heckman2000}). 
 However, the outflow velocity in BOSS-EUVLG1, 573 km s$^{-1}$, stands out against the measured weak dependence of the outflowing velocities as a function of the stellar mass. Velocities, V$_{out}$, lower than 400 km s$^{-1}$ are measured in galaxies with $\log$(M$_{stellar}$[M$_{\odot}$]) of 10.0 (lower right panel of Figure \ref{fig:results}). In summary, the observed properties of the ionized outflowing gas in BOSS-EUVLG1 appear to be a consequence of its extremely high sSFR, that is, a very massive, young, and compact starburst in a galaxy with relatively low stellar mass. On the one hand, the V$_{out}$ follows the functional dependence observed in star-forming galaxies at low and intermediate redshifts. On the other hand, the amount of outflowing material, as identified by its low F$_B$/F$_N$, appears to be consistent with the ratio expected by its low stellar mass.
 
 The blueshifts identified in the UV ISM absorption lines of BOSS-EUVLG1 are also indicative of large-scale galactic outflows, similar to that observed in many other star-forming galaxies at low \citep[e.g.,][]{Heckman2000, rubin2014} and high redshift \citep[e.g.,][]{Shapley03, Steidel+10}.  The right panel of Figure \ref{fig:HalphaUV} shows a comparison between the blueshift of the low-ionization ISM absorption lines (relative to the systemic redshift) and the UV absolute magnitude $M_{\rm UV}$ of BOSS-EUVLG1 and other UV-luminous star-forming galaxies at $2<z<3$ (most of them strongly magnified and where $\Delta v$ can be measured on an individual basis \citep[][]{pettini2000, quider2009, quider2010, des2010, patricio2016, marques2018, marques2020}).  This figure clearly shows a trend among $\Delta v_{low-ion}$ and $M_{\rm UV}$,  indicating that UV-luminous star-forming galaxies, such as BOSS-EUVLG1, experience outflows caused by supernova explosions and stellar winds, with the peak velocity of these outflows increasing with the absolute UV luminosity of the galaxy.

 A more direct comparison of other relevant physical quantities of the ionized outflow, such as the rates of outflowing mass, kinetic energy, and momentum, as well as the mass-loading factors, are difficult to establish. These quantities have a strong dependence on the value adopted for the electron density and size of the H$\alpha$ emitting nebulae. In some samples, such as the MOSDEF survey of z $\sim$ 2 of star-forming galaxies \citep{Freeman19}, a fixed radius of 3 kpc and an electron density of 50 cm$^{-3}$ is adopted for the outflowing material. Other studies of outflows in mean sequence star-forming galaxies at z $\sim$ 1 \citep{Swinbank19} adopt a radius of 10 kpc and a electron density of 75 cm$^{-3}$ for the composite spectra of different galaxies of the sample. Finally, the recently completed KMOS-3D survey of $z \sim 0.6 - 2.7$ \citep{Forster-Schreiber+19} derived an electron density of 380 cm$^{-3}$ from the stack spectra of the star-forming driven outflows. The electron densities measured or adopted in all these intermediate redshift samples are factors of $\sim$ 4 to 30 lower than the value (1500 cm$^{-3}$) measured in BOSS-EUVLG1, while the sizes are consistently larger by factors of 3-10, making  any comparison between the ionized outflow in BOSS-EUVLG1 with other galaxy samples rather uncertain in absolute terms. However, if the differences in the size and electron density were real, these would indicate that the outflow in BOSS-EUVLG1 is much more compact and involves a much denser gas than in mean sequence star-forming galaxies at intermediate redshifts. Accurate measurements in both IR-luminous and UV-luminous star-forming are needed to establish whether or not these two population of galaxies have intrinsically different physical properties in their outflows.

\subsection{The multi-phase nature of outflows}\label{multiphase}

The outflows predicted in cosmological simulations of high-z galaxies are extended over regions of several kpc in size (effective radius of 1 to 7 kpc) but with maximal velocities of less than 200 km~s$^{-1}$  \citep{Ceverino+16}, that is, a factor 2-3 smaller than the V$_{out}$ measured in BOSS-EUVLG1. However, according to most recent high angular resolution (100-200 pc) hydrodynamical simulations (TNG50; \citealt{Nelson+19}), outflowing maximal velocities tend to increase with the stellar mass of the galaxy, such that V$_{max}$ of 500-600 km s$^{-1}$ are expected in z$\sim$ 2-3 galaxies with stellar masses of about 10$^{10}$ M$_{\odot}$. According to these simulations, the mass-loading factor would be about 4 for a typical 10$^{10}$ M$_{\odot}$ galaxy at z$\sim$ 2, namely, a factor of about 100 larger than the value derived for BOSS-EUVLG1 ($\eta$= 0.05$\pm$0.03). This apparent discrepancy could be explained by the multi-phase nature of the outflows in galaxies. The H$\alpha$ line used in this study traces only the warm 10$^4$~K optical emitting gas, while it is known that some outflows carry also relevant amounts of mass, momentum, and energy in their molecular phase \citep{Veilleux+20}, and in the diffuse hot X-ray emitting gas (e.g. M82; \citealt{Strickland+09,Lopez+20}; also \citealt{Heckman+17} for a review). In fact, according to the TNG50 simulations, the ouflowing mass rate for galaxies with masses of 3 $\times$ 10$^{10}$ M$_{\odot}$, or less, is dominated by the diffuse X-ray emitting gas with the peak of the outflowing mass at temperatures above 10$^6$ K \citep{Nelson+19}. The X-ray emitting mass ouflowing rate in these galaxies is predicted to be at least a factor of 10 higher than that of the warm emitting gas \citep{Nelson+19}. It could well be that the H$\alpha$ outflow identified in BOSS-EUVLG1 could be tracing only the warm, lower velocity  phase of a much powerful outflow with a total mass-loading factor (i.e., all velocities and gas phases) close to one or even higher \citep{Nelson+19} and, therefore, relevant for the quenching or temporal delay of the star formation in these galaxies.   

\subsection{Prospects for detecting X-ray emitting outflowing gas with ATHENA}\label{ATHENA}

ATHENA, the future European Space Agency X-ray satellite, could be able to detect the X-ray emission associated with the massive starburst at the redshift of BOSS-EUVLG1. One of its instruments, the Wide Field Imager (WFI, \citealt{Meidinger+18}), will be able to detect sources with a flux of 1 $\times$ 10$^{-16}$ erg s$^{-1}$ cm$^{-2}$ in the SOFT X-ray (0.5 to 2 keV) with a signal-to-noise ratio (S/N) ratio of 5$\sigma$ in 10 hours (Athena Community Office, private communication). At the redshift of BOSS-EUVLG1, this limit corresponds to a luminosity L$_{\rm X-SOFT}$ = 5 $\times$ 10$^{42}$ erg s$^{-1}$. Well-known nearby starbursts such as M82 and NGC3079 have extended X-ray emitting halos of several kpc \citep{Strickland+04} with L$_{\rm X-SOFT}$ of 4.1 $\times$ 10$^{39}$ erg s$^{-1}$ and 8.3 $\times$ 10$^{39}$ erg s$^{-1}$, and total integrated (i.e., halo plus disk and nucleus) L$_{\rm X-SOFT}$ of 4.3 $\times$ 10$^{40}$ erg s$^{-1}$ and 3.2 $\times$ 10$^{40}$ erg s$^{-1}$, respectively. These galaxies are forming stars at a rate of 8-9 M$_{\odot}$ yr$^{-1}$ \citep{Strickland+04}, about hundred times less than in BOSS-EUVLG1. Therefore, if the efficiency in the conversion of the radiative and mechanical energy of the starburst into heating of the interstellar medium in BOSS-EUVLG1 is similar to that of M82 and NGC3079, a L$_{\rm X-SOFT}$ of 3-4 $\times$ 10$^{42}$ erg s$^{-1}$ for the integrated emission would be expected, and therefore could be detected with ATHENA. This luminosity has to be considered as a lower limit to the total X-ray emission if the ratio of the H$\alpha$ to SOFT (0.3$-2$ keV) X-ray luminosities (L$_{H\alpha}$/L$_{\rm X-SOFT}$) for the extended emission in starbursts \citep{Colina+04} is taken into account. The L$_{H\alpha}$/L$_{\rm soft-X}$ ratio covers a wide range of values from about 0.25 (NGC253), to 0.5 (NGC3079, Arp220), and 6 (M82). Applying similar ratios to BOSS-EUVLG1, a total soft X-ray luminosity in the 8 to 180 $\times$ 10$^{42}$ erg s$^{-1}$ would be expected for the outflowing ionized gas with an H$\alpha$  luminosity of 3.1 $\times$ 10$^{43}$ erg s$^{-1}$. However, the imaging capabilities of the WFI instrument precludes any direct, spatially resolved, detection of the extended emission associated with the outflowing material. WFI  will be providing X-ray images over a 40 x 40 arcmin$^2$ field-of-view with an angular resolution (FWHM) of about 5 arcsec \citep{Meidinger+18}, that is, about 40 kpc at the redshift of BOSS-EUVLG1. Our H$\alpha$ spectroscopic observations place an upper limit of a few kpc to the size of the extended emission. In addition, X-ray emitting halos is nearby starbursts have sizes of a few kpc (M82, NGC253, NGC3079) to about 25 kpc (Arp220). In summary, the detection with a high significance of the X-ray emission in starburst galaxies similar to BOSS-EUVLG1 at redshifts of 2-3 will be possible in the future with the {\it ATHENA} satellite, but the X-ray emitting halos associated with the starburst-generated outflowing material will not be spatially resolved. 


\section{Summary and Outlook}\label{Sect:summary}
We report the detection of a massive and fast ionized gas outflow in BOSS-EUVLG1, the most UV luminous star-forming galaxy detected so far in the Universe. BOSS-EUVLG1 is the brightest member of a new class of galaxies, the Extremely UV Luminous Galaxies (EUVLGs), with  (unobscured) star formation rates above 450 M$_{\odot}$ yr$^{-1}$, recently detected in the BOSS survey. The H$\alpha$ line in emission has been used to characterized the ionized outflow of BOSS-EUVLG1. H$\alpha$ presents a broad component blueshifted by -139 $\pm$ 87 km s$^{-1}$ and with a width (FWHM) of 511 $\pm$ 145 km s$^{-1}$. The presence of the outflow in BOSS-EUVLG1 is also supported by blueshifted UV ISM absorption lines. The H$\alpha$ outflow is described by a broad-to-narrow component flux ratio (F$_{B}$/F$_{N}$) of 0.34$\pm$0.12, an outflowing velocity (V$_{out}$) of 573 $\pm$ 151 km s$^{-1}$, a total mass, $\log(M_{out}[M_{\odot}]),$ of 7.94 $\pm$ 0.15,  and an outflowing mass rate ($\dot{M}_{out}$) of 44 $\pm$ 20 M$_{\odot}$ yr$^{-1}$. The velocity of the outflow in BOSS-EUVLG1 follows the expected value in the V$_{out}$-SFR diagram  predicted from the relation derived for low- and intermediate-z star-forming galaxies. The fraction of the flux carried out by the outflow is within the value expected for its stellar mass in the F$_B$/F$_N$ ratio to stellar mass diagram. 

The momentum and energy involved in the ionized gas outflow do not require the presence of an additional energy source other than the starburst itself, such as an AGN. The velocity of the outflow (V$_{out}$) is such (573 km s$^{-1}$) that only a small fraction ($\leq$ 15\%) of the outflowing mass would be able to escape and enrich the surrounding circumgalactic medium. The mass-loading factor is very low ($\eta$= 0.05 $\pm$ 0.03), indicating that, although it is massive, the ionized phase of the outflow does not play a relevant role in the quenching of the star formation. Other phases, such as the X-ray emitting gas, might be responsible for most of the outflowing mass and energy, as predicted by some recent high angular-resolution cosmological simulations. This will require confirmation in the future thanks to sensitive X-ray observations with {\it Athena}. The properties of the ionized outflow in the class of extreme UV-luminous galaxies (EUVLGs) recently identified in the BOSS survey will contribute to characterizing it further, using the H$\alpha$ emission line as a tracer.

\begin{acknowledgements}
      The  authors  thank  to the  anonymous  referee  for  useful  comments, to Natascha Forster-Schreiber for kindly sharing the data from the KMOS-3D survey used in the plots, and to Francisco Carrera and the ATHENA Community Office for providing us with the expected sensitivity for the future WFI/ATHENA instrument. This work is based on observations made with the Gran Telescopio Canarias (GTC) installed in the Spanish Observatorio del Roque de los Muchachos of the Instituto de Astrof\'{i}sica de Canarias, in the island of La Palma. This work was supported by the Spanish State Research Agency (AEI) under grants ESP2015-65597-C4-4-R, ESP2017-83197, ESP2017-86852-C4-2-R, PID2019-106280GB-I00, and MDM-2017-0737 Unidad de Excelencia ''Mar\'{i}a de Maeztu''- Centro de Astrobiolog\'{i}a (CSIC-INTA).
\end{acknowledgements}

\bibliographystyle{aa}
\bibliography{Bibliography}

\end{document}